\documentclass[11pt]{article}
\pdfoutput=1

\usepackage{lineno,hyperref}
\usepackage{color}
\usepackage{amsfonts}
\usepackage{amsmath}
\usepackage{amssymb}
\usepackage{amsthm}
\usepackage{subeqnarray}
\usepackage{lscape,graphicx,url}
\usepackage{anysize}
\usepackage{proof}
\usepackage{fancyhdr}
\usepackage{latexsym}
\usepackage{array}
\usepackage{multirow}
\usepackage{textcomp}
\usepackage{optprog}
\usepackage{mathtools}
\usepackage{natbib}

\usepackage{bm}
\usepackage{array}

\usepackage{proof}
\usepackage{paralist}
\usepackage[dvipsnames]{xcolor}
\usepackage{placeins}
\usepackage{enumitem}
\usepackage{csquotes}
\usepackage{comment}
\usepackage[none]{hyphenat}
\newcolumntype{H}{>{\setbox0=\hbox\bgroup}c<{\egroup}@{}}

\newcommand{\red}[1]{{\color{red}{#1}}}

\marginsize{3cm}{3cm}{3cm}{3cm}

\title{Subset second-order stochastic dominance for enhanced indexation with diversification enforced by sector constraints}

\author{Cristiano Arbex Valle \and John E Beasley$^2$ \and Nigel Meade$^3$}

\date{}

\begin{document}

\maketitle

\begin{center} 
{\footnotesize
$^1$Departamento de Ci\^{e}ncia da Computa\c{c}\~{a}o, \\
 Universidade Federal de Minas Gerais, \\
 Belo Horizonte, MG 31270-010, Brasil \\
arbex@dcc.ufmg.br \\ \vspace{0.3cm}
$^2$Brunel University \\ Mathematical Sciences, UK \\ john.beasley@brunel.ac.uk \\ \vspace{0.3cm}
$^3$Imperial College Business School,  Imperial College, London, UK \\ n.meade@imperial.ac.uk
}
\end{center} 

\begin{abstract}

In this paper we apply second-order stochastic dominance (SSD) to the problem of enhanced indexation with asset subset (sector) constraints. The problem we consider is how to construct a portfolio that is designed to outperform a given market index whilst having regard to the proportion of the portfolio invested in distinct market sectors. 

In our approach, subset SSD,  the portfolio associated with each sector is treated in a SSD manner. In other words in subset SSD we actively try to find sector portfolios that SSD dominate their respective sector indices, whilst
also taking into account the market index. Our subset SSD approach involves the numeric solution of a multivariate second-order stochastic dominance problem.

Computational results are given for our approach as applied to the S\&P~500 over the period $3^{\text{rd}}$ October 2018 to $29^{\text{th}}$ December 2023. This period, over 5 years, includes the Covid pandemic, which had a significant effect on stock prices. The  S\&P~500 data that we have used is made publicly available for the benefit of future researchers. Our computational results indicate that the scaled version of our subset SSD approach  outperforms the S\&P~500. Our approach also outperforms the standard SSD based approach to the problem. Our results show, that for the S\&P~500 data considered, including sector constraints improves out-of-sample performance, irrespective of the SSD approach adopted. Results are also given for Fama-French data involving 49 industry portfolios and these confirm the effectiveness of our subset SSD approach.

\end{abstract}

{\bf Keywords:} enhanced indexation, finance, optimisation, portfolio optimisation, second-order stochastic dominance

\section{Introduction}

In this paper we consider the problem of enhanced indexation with asset subset (sector) constraints. In this problem we aim to 
outperform a given market index whilst having regard to the proportion of the portfolio invested in distinct market sectors. We apply 
second-order stochastic dominance (SSD) to the problem. Computational results are given for our approach.

\sloppy We believe that the contribution of this paper to the literature  is:
\begin{itemize}
\item To present a new approach, \emph{\textbf{subset SSD}}, for the problem of enhanced indexation with asset subset (sector) constraints. This involves two stages:
 \begin{compactenum}
\item Use of an innovative subset SSD optimisation program which decides the proportion of the investment associated with each asset subset, whilst also deciding the 
investment associated with assets in each subset. \emph{\textbf{This subset SSD optimisation program explicitly takes into account the performance of an index for each asset subset
as well as the performance of the market index.}}
\item Given the proportional investment in each subset, utilising degeneracy in an attempt to improve upon 
the investment in individual assets by finding asset subset SSD portfolios.
\end{compactenum}
\item To demonstrate computationally,
using data that we make publicly available:  
 \begin{compactitem}
\item that our subset SSD approach  outperforms both the  S\&P~500 and the standard SSD approach to the problem
\item that including sector constraints improves out-of-sample performance, irrespective of the SSD approach adopted
\end{compactitem}
\end{itemize}

The structure of this paper is as follows. In Section~\ref{sec2} we review the relevant literature as to second-order stochastic 
dominance. 
In Section~\ref{sec4} we present our subset SSD approach  when we have sector (asset subset) constraints present that constrain 
investment in a number of different subsets of assets. 
We also give the standard SSD approach to the problem.
In Section~\ref{sec5} we present the computational results obtained 
when our subset SSD approach is applied to the S\&P~500. 
In addition we also apply our approach to Fama-French data involving 49 industry portfolios. 
In Section~\ref{sec6} we present our conclusions.

\section{Literature review}
\label{sec2}

\sloppy In this section we review the literature relating to stochastic dominance and second-order stochastic dominance that has particular
relevance to the work 
presented in this paper. The reader should note here that this section is not an extensive review of the literature relating to 
these two topics. That would be a mammoth task far outside the scope of this paper. For example a recent search using 
Web of Science (\url{http://www.webofscience.com}) listed over 2,900 papers referring to \enquote{stochastic dominance} and over 330 papers referring 
to \enquote{second-order stochastic dominance}.

\sloppy The importance of stochastic dominance (SD) within financial portfolio selection has been recognised for decades~\citep{hadar69, bawa1975, levy1992}. For two 
random variables $X$ and $Y$ it is well known that $X$ dominates $Y$ under first-order stochastic
 dominance (FSD, $ X \succeq_{_{FSD}} Y$) if and only if it is preferrable over any monotonic increasing utility function. Likewise, $X$ dominates $Y$ under
 second-order stochastic dominance (SSD, $ X \succeq_{_{SSD}} Y$) if and only if it is preferrable over any increasing and strictly concave (risk-averse) utility function~\cite{hadar69, hanoch69}.

For many years however SD was primarily a theoretical framework in terms of financial portfolio optimisation. This was
 due to the perceived computational difficulties associated with finding SD-efficient portfolios. In the past twenty years, however, there has been a shift towards applying SD (especially SSD) principles in practice, with several optimisation approaches having been proposed for finding portfolios that are either SSD-efficient (with regards to a
specified set of feasible portfolios) or SSD-dominating (with regards to a benchmark).

\cite{ogryczak2002} identified several risk measures that can be employed in mean-risk ($\mu_x, r_X$) decision models that are consistent with the SSD relation in the sense that $X \succeq_{_{SSD}} Y$ implies that $\mu_X \geq \mu_Y$ and $r_X \leq r_Y$. These measures include tail value-at-risk, tail Gini mean difference and weighted mean deviation from a quantile. The authors presented stochastic linear programming formulations for these models whose optimal solutions are guaranteed to be SSD-efficient.

~\cite{kuosmanen2004, kuosmanen2001}
developed the first SSD efficiency tests based on mathematical programming. 
Their formulation finds, if it exists,  
the portfolio with the highest in-sample mean that dominates a benchmark in the SSD sense.
\cite{post2003} developed linear programming models for testing if a given portfolio is SSD-efficient with respect to all possible portfolios given a set of assets. 

\cite{dentcheva2006, dentcheva2003} first combine the available assets to produce a reference (or benchmark) distribution, and then compute a portfolio which SSD-dominates the benchmark. They used the lower partial moment of order one to develop the SSD ranking concerning the benchmark portfolio. Their work has been the basis of several later papers in the literature, as referenced below.

\cite{roman2006} introduced a multi-objective optimisation model to find a portfolio that achieves SSD dominance over a benchmark. If no such portfolio exists they find the portfolio whose return distribution comes closest to the benchmark. 
\cite{luedtke2008} presented a model that generalises that of~\cite{kuosmanen2004} which includes FSD constraints based on a cutting-plane formulation for problems with integrated chance constraints. Their model involves integer variables, but relaxing integrality yields a formulation with SSD constraints. 

\cite{fabian2011a,fabian2011b} introduced a cutting plane reformulation of~\cite{roman2006} which generalises~\cite{dentcheva2006}. The authors replaced the multi-objective nature of the problem by maximising the minimum value in the SSD relation with regards to a benchmark. 
\cite{roman2013} applied the SSD cutting plane formulation in an enhanced indexation setting.~\cite{valle2017} added exogenous constraints and 
reformulated the problem as an integer linear program, for which a branch-and-cut algorithm was developed.

\cite{kopa2015, post2013} introduced a more generalised efficiency test which allows for unequal probabilities and higher orders. In the case of inefficiency
their dual 
model finds a dominating portfolio. If the portfolio being tested is a benchmark, this dual model can be seen as equivalent to a model for
enhanced indexation. 

The set of SSD efficient portfolios is generally very large, and investors need to decide how to select a portfolio in which to invest
from within  this set. The formulation from~\cite{post2013} may be used to find different SSD-efficient portfolios depending on how some parameters are specified.~\cite{hodder2015} proposed ways to assign values to these parameters with the goal of helping investors select a single portfolio out of the efficient set.

\cite{bruni2017, bruni2012} developed an alternative approach for SD-based enhanced indexation. They proposed
 a criterion called ``cumulative zero-order stochastic $\epsilon$-dominance'' (CZS$\epsilon$D). Zero-order SD happens when all returns from a given portfolio are superior to all returns from an alternative portfolio. 

\cite{sharma2017} introduced a relaxed-SSD formulation for enhanced indexation. The SSD constraints are relaxed by adding under/overachievement
where  SSD violation is controlled by setting an appropriate upper bound related to the total underachievement. The concept of relaxed-SSD was first introduced by~\cite{lizyayev2012}. 

\cite{sharma2017b} proposed a SSD-based approach for producing sector portfolios. 
For each sector their model seeks a SSD portfolio that dominates the corresponding sector index whilst focusing on 
a number of financial ratios such as net earnings/total assets.
These sector portfolios are then combined using another model that optimises 
their mean return subject to being  (if possible) SSD-dominating with respect to the main market index. If SSD dominance cannot be 
achieved
they 
relax the dominance constraints in their models.

\cite{post18} presented an approach based on probability estimation using non-parametric empirical likelihood followed by 
generation of a SSD portfolio. Probability estimation involved the solution of a nonlinear program. 
\cite{liesio20} extended SSD by relaxing the assumption that the state (scenario) probabilities were well-specified (typically equally-likely). 
In their approach they identify portfolios that dominate the benchmark for any state probabilities in a given set.

\cite{liu2021} showed that FSD and SSD may not be sufficient to discriminate between
multiple dominating portfolios with regards to a benchmark. They proposed a new criterion called Interval-based SD (ISD) in 
which different SD orders are applied to different parts of the support of the return distribution. 
\cite{malavasi21} compared the performance of SSD portfolios with efficient portfolios derived using the standard mean-variance approach 
of~\cite{mark52}. They also focused on the performance of the global minimum variance portfolio as compared with portfolios 
that are stochastically dominant to this minimum variance portfolio.

\cite{sehgal2021} presented a robust version of the SSD-formulation of~\cite{dentcheva2006}. Robustness is introduced by varying asset returns, and the model is developed as the deterministic equivalent of a stochastic programming formulation.~\cite{goel2021} also generalised~\cite{dentcheva2006} by considering the ``utility improvement'' in portfolio returns instead of the returns themselves. The authors proposed replacing the portfolio and benchmark returns by their respective deviations in the SSD constraints.
\cite{cesarone2022} compared the formulations of~\cite{roman2013} and~\cite{kopa2015} with skewed benchmarks obtained by using
 the reshaping method of~\cite{valle2017}. They found that SSD portfolios that dominate the skewed benchmark generally
perform better out-of-sample. 

\cite{liesio23} considered the problem of generating an efficient  frontier using stochastic dominance. They presented an approach based on Pareto optimal solutions of a multiple objective optimisation problem.
\cite{xu24} presented an approach aiming to identify robust portfolios whose in-sample SSD dominance is likely to 
hold out-of-sample. 
\cite{cesarone2025} presented an approach based upon ordered weighted average in conjunction with stochastic dominance. 
The weighted average is based on the sum of $k$ maximal CVaRs (where CVaR is 
Conditional Value at Risk).

In the work discussed above we are typically aiming to find a portfolio that is second-order stochastic dominant with respect to a 
benchmark. In work of this kind we  are dealing with 
\textit{\textbf{univariate second-order stochastic dominance}}  in that we have a 
single variate (e.g.~return) associated with the chosen portfolio, a single variate (e.g.~index return) associated with the benchmark.

The work dealt with in this paper deals with 
\textit{\textbf{multivariate second-order stochastic dominance}}. Here we are choosing a single portfolio, 
but within that portfolio we have different asset subsets with different characteristics, and for each subset 
 we have a benchmark  to compare against. 

As a simple example within a single portfolio made up from stocks (equities) we might have some investment in  large capitalisation (large-cap) companies and some 
investment in 
small capitalisation (small-cap) companies. Here we compare the  returns from these two asset subsets (so we have two variates, 
returns for each subset) against two benchmark equity indexes (so again two variates, one being an 
index of returns for all large-cap stocks, the other being an index of returns for all small-cap stocks).
Readers interested in this broader field of multivariate stochastic 
dominance are referred to~\cite{arvanitis24, kopa23, arvanitis21, armbruster15, dentcheva09, muller02}.

\section{Subset SSD}
\label{sec4}

In this section we present our subset SSD approach for enhanced indexation with sector constraints. We first discuss 
the motivation behind our approach and give an overview of it. We then give the mathematical details associated with our approach.
We indicate how, given the proportional investment in each asset subset, we can 
potentially improve upon the investment in asset subsets by finding individual subset SSD portfolios.

For brevity here we will assume that
the reader has familiarity with SSD. 
Readers unfamiliar with this topic are referred to~\cite{fabian2011a, fabian2011b, roman2006}.

\subsection{Motivation and overview}

In the literature survey above a common context was to  have a single set of financial assets and 
to seek a portfolio chosen from these assets that, in a SSD sense, outperforms  (if possible) in terms of the return 
achieved the return from an associated index. In other words a univariate second-order stochastic dominance approach.

In subset SSD we generalise this approach to the case where it is possible to subdivide the entire set of assets into 
individual subsets, each with differing characteristics. In other words a multivariate second-order stochastic dominance approach.

\textit{\textbf{The principal motivation behind our subset SSD approach can be summarised in a single word - diversification}}. The importance of diversification in 
choosing a financial portfolio has long been recognised. 
If we are to ensure that a SSD portfolio is suitably diversified one approach is to classify the underlying assets into distinct subsets 
according to common characteristics, and introduce constraints restricting the investment in each such subset. 
\textbf{\textit{But whilst in standard SSD we compare the portfolio chosen against a market index in subset SSD
 we also compare each asset subset against a market index for that subset.}}

How to classify assets into subsets can be easily envisaged e.g.~based on market capitalisation,  market sectors, 
momentum characteristics or any other economic metric. 
\cite{post18} note that in terms of adopting an equity industry momentum strategy diversification is relevant as a concentrated position 
in just one or two top-performing industries would be riskier than most investors would be willing to tolerate.

As an overview of our subset SSD approach we regard the chosen overall portfolio  as made 
up of a set of asset subset portfolios, with each asset subset portfolio being made up of  investment in assets drawn from that subset.
So  in subset SSD we aim to choose a portfolio in which each of the 
asset subset portfolios outperform in a SSD sense (if possible) an associated asset subset index. However we also consider the performance of the 
overall portfolio  as compared with the overall market index.
This is accomplished in two stages:
\begin{compactenum}
\item Use of an innovative subset SSD optimisation program which decides the proportion of the investment associated with each asset subset, whilst also deciding the 
investment associated with assets in each subset. This subset SSD optimisation program explicitly takes into account the performance of an index for each asset subset
as well as the performance of the market index.
\item Given the proportional investment in each subset, utilising degeneracy in an attempt to improve upon 
the investment in individual assets by finding asset subset SSD portfolios.
\end{compactenum}

\subsection{Subset SSD optimisation program}
In this section we give our innovative subset SSD optimisation program. Let:
\begin{compactitem}
\item $N$  be number of assets available for  investment
\item $K$ be the number of asset subsets
\item  $N^k$ be the assets in asset subset $k$ where $\cup^K_{k=1} N^k= [1,...,N]$. In our approach 
we do not assume that the asset subsets are disjoint, in other words a single asset can be 
in two or more subsets. For ease of presentation we let asset subset zero represent
the set of all assets, so $N^0 = [1,\ldots,N]$.
\item $S$ be  number of scenarios, where the scenarios are assumed to be equiprobable
\item $r_{is}$ be the return of asset $i$ in scenario $s$
\item $R_s^P$ be the return associated with a given asset portfolio $P$ in scenario $s$
\end{compactitem}

\noindent SSD makes use of the left tail of the cumulative return distribution associated with $[R_1^P, R_2^P,\ldots,R_S^P]$
weighted by the constant $(1/S)$ factor. So:
\begin{equation}
 \text{Tail}^L_{\frac{s}{S}}(P) = \frac{1}{S} \text{(sum of the $s$ smallest portfolio returns in $[R_1^P, R_2^P,\ldots,R_S^P]$)}
\label{jebt1}
\end{equation}

We need for each asset subset $N^k$ an underlying index in order to create an appropriate SSD formulation. Such an index may be publicly available. If not,
 one can easily be produced using weights associated with any index that includes these assets. 
Let $I^k$ represent the returns on the index associated with asset subset $k$. Then define 
$(\hat{\tau}^k_1, \ldots, \hat{\tau}^k_S) = \big( \text{Tail}^L_{\frac{1}{S}} I^k, \ldots, \text{Tail}^L_{\frac{S}{S}} I^k \big)$.
The variables associated with our subset SSD program are:
\begin{compactitem}
\item  $\mathcal{V}_s^k$ the tail difference between the chosen portfolio and the index portfolio associated with asset subset $k$
\item  $w_i \geq 0$  the proportion of the portfolio invested in asset $i$ 
\item $W^k \geq 0$ the proportion of the portfolio invested in asset subset $k$
\end{compactitem}
The constraints of the subset SSD optimisation program are:
\begin{equation}
\mathcal{V}_s^k \leq \frac{1}{S} \sum_{j \in \mathcal{J}} \sum_{i \in N^k} r_{ij} w_i /W^k- \hat{\tau}_s^k~~~~\forall \mathcal{J} \subseteq \{1, ..., S\},~|\mathcal{J}| = s,~s=1,\ldots,S,~k=0,\ldots,K
\label{exjebt5s}
\end{equation}
\begin{equation}
 W^k = \sum_{i \in N^k} w_i~~~~k=0,\ldots,K
\label{exjebt6s}
\end{equation}
\begin{equation}
\delta^L_k \leq W^k \leq \delta^U_k~~~~k=1,\ldots,K
\label{eqjebdelta}
\end{equation}
\begin{equation}
 \sum_{i=1}^N w_i =1
\label{exjebt6as}
\end{equation}
\begin{equation}
w_i \geq 0~~~~i=1,\ldots,N
\label{exjebt7s}
\end{equation}
\begin{equation}
\mathcal{V}_s^k \in\mathbb{R}~~~~s=1,\ldots,S,~k=0,\ldots,K
\label{exjebt8s}
\end{equation}

Equation~(\ref{exjebt5s}) is the tail difference for each subset $k$. In this equation the summation in the numerator of the first 
term on the right-hand side of the inequality is the return from the investment in assets associated with subset $k$. 
As the sum of the weights (over assets $i \in N^k$) may not equal one
 we have to divide this summation by the $W^k$ factor before subtracting the
tail difference  $ \hat{\tau}_s^k$ associated with subset $k$.

Equation~(\ref{exjebt5s}) defines the  $s$ smallest portfolio returns in the $S$ scenarios using a combinatorial number of constraints.
Note here that in this equation we have the tail difference for all of the $K$ asset subsets as well as 
the tail difference for the market index (asset subset zero).

Equation~(\ref{exjebt6s}) defines the subset proportion  based on the sum of the proportions of the total wealth invested in the assets in the subset. 
Equation~(\ref{eqjebdelta}) ensures that the proportion of the total investment in subset $k$ lies between   $\delta^L_k$ and $\delta^U_k$ where these are
the user defined lower and upper limits on the proportion of the portfolio invested in subset $k$.
Equation~(\ref{exjebt6as}) ensures that all of our wealth is invested in assets.
Equation~(\ref{exjebt7s}) is the non-negativity constraint 
(so no short-selling) and Equation~(\ref{exjebt8s}) ensures that the tail differences $\mathcal{V}^k_s$ can be positive or negative.

Equation~(\ref{exjebt5s}) is nonlinear but by assuming that $W^k > 0~k=1,\ldots,K$ (which we can ensure if we wish by adding constraints $W^k \geq \epsilon~k=1,\ldots,K$, where $\epsilon{>}0$ and small) we can linearise it to:
\begin{equation}
W^k \mathcal{V}_s^k \leq \frac{1}{S} \sum_{j \in \mathcal{J}} \sum_{i \in N^k} r_{ij} w_i – W^k \hat{\tau}_s^k~~~~\forall \mathcal{J} \subseteq \{1, ..., S\},~|\mathcal{J}| = s,~s=1,\ldots,S,~k=0,\ldots,K
\label{exjebt5slin}
\end{equation}
Here the $ W^k \mathcal{V}_s^k$  term is nonlinear, but can be interpreted as the \emph{\textbf{proportion weighted tail difference}} associated with set $k$. 

In our subset SSD optimisation program we focus on maximising the minimum value of this proportion weighted tail difference.
Replace the nonlinear proportion weighted tail difference term $ W^k \mathcal{V}_s^k$ 
in Equation~(\ref{exjebt5slin}) by a single term, say  $\mathcal{Z}_s^k \in\mathbb{R}$, and adopt an objective function of the form:
\begin{equation}
\mbox{maximise}~\mathcal{V}
\label{exjebt4sz}
\end{equation}
where we optimise this objective subject to:
\begin{equation}
 \mathcal{Z}_s^k \leq \frac{1}{S} \sum_{j \in \mathcal{J}} \sum_{i \in N^k} r_{ij} w_i – W^k \hat{\tau}_s^k~~~~\forall \mathcal{J} \subseteq \{1, ..., S\},~|\mathcal{J}| = s,~s=1,\ldots,S,~k=0,\ldots,K
\label{exjebt5slina}
\end{equation}
\begin{equation}
\beta_s \mathcal{V} \leq  \mathcal{Z}_s^k  ~~~~s=1,\ldots,S,~k=0,\ldots,K
\label{exjebt5slina1}
\end{equation}
\begin{equation}
 W^k = \sum_{i \in N^k} w_i~~~~k=0,\ldots,K
\label{exjebt6s9}
\end{equation}
\begin{equation}
\delta^L_k \leq W^k \leq \delta^U_k~~~~k=1,\ldots,K
\label{eqjebdelta9}
\end{equation}
\begin{equation}
 \sum_{i=1}^N w_i =1
\label{exjebt6sa9}
\end{equation}
\begin{equation}
w_i \geq 0~~~~i=1,\ldots,N
\label{exjebt7s9}
\end{equation}
\begin{equation}
 \mathcal{V} \in\mathbb{R}
\label{jebt8as}
\end{equation}
\begin{equation}
\mathcal{Z}_s^k \in\mathbb{R}~~~~s=1,\ldots,S,~k=0,\ldots,K
\label{zreal}
\end{equation}
In Equation~(\ref{exjebt5slina1}) $\beta_s$ is the scaling factor where $\beta_s{=}1$ for no scaling and $\beta_s{=}s/S$ for scaled tails.
Scaling was proposed by~\cite{fabian2011b} to ensure that more importance is given to the returns in the right tails of the distribution. 
Equations~(\ref{exjebt6s9})-(\ref{exjebt7s9}) 
are  as
 Equations~(\ref{exjebt6s})-(\ref{exjebt7s}) from before.  
 Equations~(\ref{jebt8as}),(\ref{zreal}) ensure that the variables involved  can be positive or negative.

With respect to a  minor technical issue there is no need to include the nonlinear constraint that 
$ \mathcal{Z}_s^k = W^k \mathcal{V}_s^k$ here, since given values for $ \mathcal{Z}_s^k$ and  $W^k$ from the solution to 
this optimisation program a valid value for $\mathcal{V}_s^k \in\mathbb{R}$ is given by $ \mathcal{V}_s^k  =\mathcal{Z}_s^k /W^k$.

The above subset SSD formulation,
Equations~(\ref{exjebt4sz})-(\ref{zreal}), is a linear program for which, as the reader may be aware,
there exist sophisticated and powerful solution packages.
Computationally the difficulty with solving this linear program 
 to proven optimality relates to the combinatorial number of constraints associated with Equation~(\ref{exjebt5slina}). However we can easily adapt the cutting plane procedure given previously by~\cite{fabian2011a} in order to deal with this difficulty. For completeness  we set  out  this cutting plane procedure in full in~\ref{append}.

\subsection{Individual subset SSD portfolios}
\label{sec:gamma}

The subset SSD optimisation program given above, Equations~(\ref{exjebt4sz})-(\ref{zreal}), 
can be seen as deciding the appropriate investment proportions $W^k$ for each asset subset $k$. Because of degeneracy, alternative optimal solutions, solving
the subset SSD program  does not necessarily return a SSD portfolio for each and every asset subset.

To illustrate this suppose that we have just two disjoint subsets and are considering the unscaled subset SSD program (so $\beta_s=1$ in Equation~(\ref{exjebt5slina1})). Suppose that in the solution to the subset SSD optimisation program the subset proportions are $W^1_{opt}$ and  $W^2_{opt}$. Suppose that
the maximum value of $\mathcal{V}$ is equal to the minimum value of $\mathcal{Z}_s^1~s=1,\ldots,S$, so associated with subset 1, but that the maximum value of $\mathcal{V}$ is strictly less than the minimum value of $\mathcal{Z}_s^2~s=1,\ldots,S$. If this is the case then there may well be alternative optimal solutions, \emph{\textbf{but with the same asset subset investment proportions}}, for which the minimum value of $\mathcal{Z}_s^2~s=1,\ldots,S$ can be increased. If this minimum value can be increased, and since $\mathcal{Z}_s^2 = W^2_{opt}  \mathcal{V}_s^2$, where $W^2_{opt}$ the proportion of the investment in asset subset 2 has been fixed, then this implies 
that for subset 2 we can increase the minimum value of $\mathcal{V}_s^2$, which is what we seek to do if we are trying to find a portfolio for 
subset 2 which is a SSD portfolio for that subset.

This example illustrates that we need to proceed to resolve degeneracy after solving the subset SSD optimisation program. 
Let the optimal values for the proportion $W^k$ of the portfolio invested in subset $k$ be $W^k_{opt}$.
Since, in general,  asset subsets may not be disjoint, we need to jointly consider  any asset subsets that have one or more assets in common. In order to do this we  repetitively identify a collection of asset subsets $\Gamma$ in the following manner:
\begin{enumerate}
\item $\Gamma^{all} \leftarrow [k~|~k=1,\ldots,K]$
\item While $\Gamma^{all}$ not empty do:
\begin{enumerate}
 \item [3.]  Set   $\Gamma\leftarrow \emptyset$. Choose a random asset subset $k \in \Gamma^{all}$ and set $\Gamma \leftarrow \Gamma \cup [k]$,  
$\Gamma^{all} \leftarrow \Gamma^{all} \setminus [k]$ 
 \item[4.]  Repeat until no further action can be taken: if any asset subset $k \in \Gamma^{all}$ has assets in common with one or more asset subsets in
$\Gamma$, 
then set $\Gamma \leftarrow \Gamma \cup [k]$,  $\Gamma^{all} \leftarrow \Gamma^{all} \setminus [k]$ 
\item [5.] Deal with $\Gamma$ using the optimisation program given below.
\end{enumerate}
\end{enumerate}
\noindent In the first step here we initialise $\Gamma^{all}$ to be all asset subsets. The second step ensures that we continue
steps 3-5 until $\Gamma^{all}$ is empty.
In the third step we chose a random asset subset in $\Gamma^{all}$, updating $\Gamma$ and
$\Gamma^{all}$  appropriately. In the fourth step we repetitively add to $\Gamma$ any asset subsets remaining in $\Gamma^{all}$ which have assets in common with a subset currently in $\Gamma$. Once this step terminates $\Gamma$ will be a maximal asset subset where the members of $\Gamma$ have assets in common and we deal with it using the optimisation program given below.

In summary the above procedure repetitively identifies maximal asset subsets $\Gamma$ which have assets in common. If all asset subsets 
are disjoint then it will repetitively identify just a single asset subset.
To adjust the proportions associated with the assets involved in $\Gamma$ we solve the following program:
\begin{equation}
\mbox{maximise}~\mathcal{V}
\label{ajeb1}
\end{equation}
subject to 
\begin{equation}
\mathcal{V}_s^k \leq \frac{1}{S} \sum_{j \in \mathcal{J}} \sum_{i \in N^k} r_{ij} w_i /W^k_{opt}- \hat{\tau}_s^k~~~~\forall \mathcal{J} \subseteq \{1, ..., S\},~|\mathcal{J}| = s,~s=1,\ldots,S,~\forall k \in \Gamma
\label{ajeb2}
\end{equation}
\begin{equation}
\beta_s \mathcal{V} \leq W^k_{opt} \mathcal{V}_s^k  ~~~~s=1,\ldots,S,~\forall k \in \Gamma
\label{ajeb2a}
\end{equation}
\begin{equation}
W^k_{opt} = \sum_{i \in N^k} w_i~~~~\forall k \in \Gamma
\label{ajeb3}
\end{equation}
\begin{equation}
w_i \geq 0~~~~\forall i \in [~j~|~j \in {N^k},~\forall k \in \Gamma]
\label{ajeb5}
\end{equation}
\begin{equation}
\mathcal{V} \in\mathbb{R}
\label{ajeb6}
\end{equation}
\begin{equation}
\mathcal{V}_s^k \in\mathbb{R}~~~~s=1,\ldots,S,~\forall k \in \Gamma
\label{ajeb7}
\end{equation}

Equation~(\ref{ajeb1}) is the standard SSD objective. 
Equation~(\ref{ajeb2}) is the same as Equation~(\ref{exjebt5s}) except that we use the subset SSD value decided for $W^k$ and restrict attention to asset
subsets $k \in \Gamma$. 
Equation~(\ref{ajeb2a}) is the standard scaled SSD equation relating $\mathcal{V}$ to $\mathcal{V}^k_s$, but restricted to just  asset
subsets $k \in \Gamma$, where the  $\mathcal{V}^k_s$ term is scaled using $W^k_{opt}$, c.f.~Equation~(\ref{exjebt5slina1}).
 Equation~(\ref{ajeb3}) ensures that asset subset $k$ retains the
 same investment proportion as decided by the subset SSD optimisation program. Equation~(\ref{ajeb5}) is as Equation~(\ref{exjebt7s}),
 but restricted to assets 
involved in the asset subsets in $\Gamma$. Equations~(\ref{ajeb6}),(\ref{ajeb7}) are the standard
 conditions for  $\mathcal{V}$  and $\mathcal{V}_s^k$ which allow them to be positive or negative. 

Note here that in the above optimisation, Equations~(\ref{ajeb1})-(\ref{ajeb7}), we only consider asset subset indices
$k=1,\ldots,K$, the market 
index ($k=0$) plays no role. This is because the market index, as well as the asset subset indices, both played a role in deciding 
the asset subset proportions $W^k_{opt}$.  Here, as these subset 
proportions are now fixed, and as we 
only seeking SSD portfolios for asset subsets, we exclude the market index from consideration.

This optimisation program, Equations~(\ref{ajeb1})-(\ref{ajeb7}), can be solved using a modified version
of the  cutting plane procedure given in~\ref{append}. 
Since the modifications needed are minor we, for space reasons, have not given that modified procedure here.

If $|\Gamma| = 1$, so we are dealing with a single asset subset $k$ , then the above optimisation program is equivalent to finding a SSD portfolio for asset subset $k$ and scaling the weights in that SSD portfolio by $W^k_{opt}$. See~\ref{append1} for a proof of this.

If  $|\Gamma| \geq 2$, so we are dealing with two or more asset subsets with assets in common, achieving a SSD portfolio for one asset subset might mean that another asset subset cannot achieve a SSD portfolio. In cases such as this we can either just use Equations~(\ref{ajeb1})-(\ref{ajeb7}) or give some explicit tradeoff between asset subsets and amend Equations~(\ref{ajeb1})-(\ref{ajeb7}) accordingly. Since the data we used had no cases with $|\Gamma| \geq 2$ we have not explored this issue further here.

\subsection{Standard SSD based approach}
\label{secalt}

Above we  presented our subset SSD approach where each individual asset subset is treated in a SSD manner. The standard
SSD based approach to the problem 
as to how to construct a portfolio that is
designed to outperform a given market index whilst having regard to the proportion of the portfolio invested in distinct asset
subsets (market sectors)
is to add 
 constraints related to asset subsets
to the SSD formulation. Using the notation established above this standard approach is:

\begin{equation}
\mbox{maximise} ~ \mathcal{V}
\label{jebt4}
\end{equation}
subject to 
\begin{equation}
\mathcal{V}_s^0  \leq \frac{1}{S} \sum_{j \in \mathcal{J}} \sum_{i = 1}^N r_{ij} w_i - \hat{\tau}_s^0~~~~\forall \mathcal{J} \subseteq \{1, ..., S\},~|\mathcal{J}| = s,~s=1,\ldots,S
\label{jebt5}
\end{equation}
\begin{equation}
\beta_s \mathcal{V}  \leq \mathcal{V}_s^0 ~~~~s=1,\ldots,S
\label{jebt5fab}
\end{equation}
\begin{equation}
 W^k = \sum_{i \in N^k} w_i~~~~k=0,\ldots,K
\label{exjebt6sa}
\end{equation}
\begin{equation}
\delta^L_k \leq W^k \leq \delta^U_k~~~~k=1,\ldots,K
\label{eqjebdeltaa}
\end{equation}
\begin{equation}
 \sum_{i=1}^N w_i =1
\label{jebt6}
\end{equation}
\begin{equation}
 w_i \geq 0~~~~i=1,\ldots,N
\label{jebt7}
\end{equation}
\begin{equation}
 \mathcal{V} \in\mathbb{R}
\label{jebt8a}
\end{equation}
\begin{equation}
 \mathcal{V}_s^0 \in\mathbb{R}~~~s=1,\ldots,S
\label{jebt8}
\end{equation}

\sloppy Equation~(\ref{jebt4}), in conjunction with Equation~(\ref{jebt5fab}),  maximises the minimum tail difference. 
Again we make  use of the scaling term $\beta_s$ in Equation~(\ref{jebt5fab}).
Equation~(\ref{jebt5}) defines the tail differences. 

Equation~(\ref{jebt6}) ensures that all of our wealth is invested in assets. Equation~(\ref{jebt7}) is the non-negativity constraint 
(so no short-selling). Equation~(\ref{jebt8a}) ensures that  $\mathcal{V}$ can be positive or negative whilst
Equation~(\ref{jebt8}) ensures that the tail differences $\mathcal{V}_s^0$ can be positive or negative.
Here we have added the subset constraints, Equations~(\ref{exjebt6sa}),(\ref{eqjebdeltaa}), to the standard SSD formulation.

Note here that this standard approach makes no use of any sector indices (unlike our subset SSD approach given above).

\section{Computational results}
\label{sec5}

In this section we first discuss the data sets that we used. These were associated with the S\&P~500 and Fama-French industry portfolios. We then go on to give results for our subset SSD approach, as well as the standard SSD based approach, on our S\&P~500 data. Following this we investigate the performance of our subset SSD approach on our S\&P~500 data when we vary sector bounds. 
 We go to investigate 
out-of-sample performance when no sector constraints are applied.
Results are then given for Fama-French data relating to 49 industry portfolios, both when the benchmark is an equally-weighted portfolio and when the benchmark is an all-share index from the Center for Research in Security Prices (CRSP). 

We used~\cite{cplex} as the
linear  programming solver, with default options. Our backtesting tool is developed in Python and all optimisation models are developed in C++. We ran all experiments on an Intel(R) Core(TM) i7-3770 CPU @ 3.90GHz with 8 cores, 8GB RAM and with Ubuntu 22.04.3 LTS as the operating system.

\subsection{Data sets}

\subsubsection{S\&P~500 data set}
We used a dataset associated with the S\&P~500, with daily stock prices  from $3^{\text{rd}}$ October 2018 until $29^{\text{th}}$
 December 2023.
This time period, over 5 years,  includes the Covid pandemic, which had a significant effect on stock prices.
Our data has been manually adjusted to account for survivorship bias - on a given date only assets that 
were part of the S\&P~500 index at that time are available to be selected for investment.

In order to define the scenarios  required by SSD we used a lookback approach that included the 
most recent 61 daily prices, which then yield 60 in-sample returns (roughly a trimester in business days). 
We have one scenario for each in-sample time period considered, where a single scenario
consists of  the return for each  asset, as well as the overall market index return and the index return for 
each asset subset.

The SSD subsets were defined by the economic sectors to which each asset belongs. There are 11 different stock market sectors 
according to the most commonly used classification system, known as the 
Global Industry Classification Standard (GICS). These sectors are 
communication services,
consumer discretionary,
consumer staples,
energy, 
financials,
healthcare,
industrials,
materials, 
real estate,
technology and 
utilities.
For each sector, the  benchmark  consisted of the corresponding time series for the S\&P sector indices, see \\
 \url{https://www.spglobal.com/spdji/en/index-family/equity/us-equity/sp-sectors/}. Table~\ref{table1} shows the S\&P~500 sector breakdown as of $9^{\text{th}}$ October 2023 together with the approximate weight of the sector with regard to the index.

\begin{table}[!ht]
\centering
{\small
\renewcommand{\tabcolsep}{1mm} 
\renewcommand{\arraystretch}{1} 
\begin{tabular}{|l|c|}
\hline
\multicolumn{1}{|c|}{Sector} & \multicolumn{1}{c|}{Approximate weight (\%)} \\
\hline
Technology               & 26.0 \\
Healthcare               & 14.5 \\
Financials               & 12.9 \\
Consumer discretionary   & 9.9 \\
Industrials              & 8.6 \\
Communication services   & 8.2 \\
Consumer staples         & 7.4 \\
Energy                   & 4.5 \\
Utilities                & 2.9 \\
Materials                & 2.6 \\
Real estate              & 2.5 \\
\hline
\end{tabular}
}
\caption{S\&P~500 sector breakdown}
\label{table1}
\end{table}

\subsubsection{Fama-French data set}
It has been common in the literature for papers dealing with SSD to make use of the Fama-French data relating to the returns on industry portfolios,  see \\
\url{https://mba.tuck.dartmouth.edu/pages/faculty/ken.french/data_library.html}. 

In the SSD work dealing with Fama-French data some papers  (e.g.~\cite{ cesarone2025,   goel2021, 
bruni2017}) use a benchmark comprising an equally-weighted portfolio, this can be easily found from the original Fama-French data. Other papers (e.g.~\cite{xu24,  post18, post2017m, hodder2015}) use a benchmark consisting of the all-share index from the Center for Research in Security Prices (CRSP), a value-weighted average of common stocks listed on the NYSE, AMEX and NASDAQ stock exchanges.  Sector indices associated with CRSP also exist, see 
\url{https://www.crsp.org/indexes/}.
As best as we are aware CRSP data is not freely available, rather access to it requires an institutional subscription and we were fortunate to have a coauthor with institutional access so that we could make use of both the CRSP all-share index and the CRSP sector indices.

\begin{figure}[!htb]
\centering
\includegraphics[width=1\textwidth]{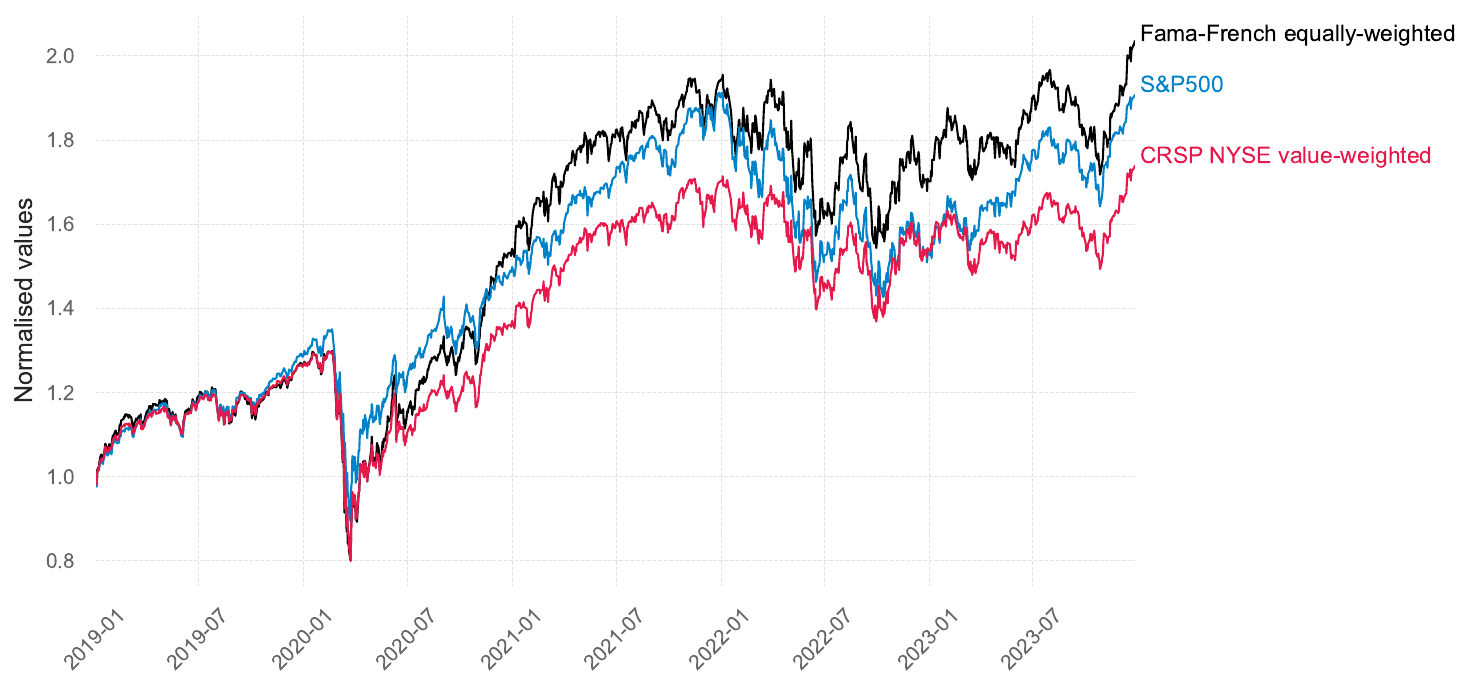}
  \caption{Index values over time}
  \label{fig1}
\end{figure}

Figure~\ref{fig1} shows the values of the three indices used in this paper over time. The effect of 
the Covid pandemic, 
which had a significant impact on stock prices in the first half of 2020, can be clearly seen there.

Both the S\&P~500 data and the Fama-French data used in this paper are publicly available for use by other researchers at:
\begin{center}
\url{https://github.com/cristianoarbex/subsetSSDData/}
\end{center}
We hope that the publicly available data set of S\&P~500 stock prices we make available, manually curated  to account for survivorship bias,  will contribute to future research, both with regard to SSD and with regard to finance portfolios more generally.

\subsection{Out-of-sample performance - subset SSD with sector bounds}
\label{sec:subsetout}

In this section we evaluate the performance of our subset SSD 
approach when compared to the S\&P~500.

As mentioned above we used an in-sample period of 61 days. 
We conducted periodic rebalancing every 21 days (roughly one month in business days). 
To illustrate our approach our first in-sample period of 61 days runs from $3^{\text{rd}}$ October 2018 until $31^{\text{st}}$ December 2018.
So using this in-sample period (with 60 return values for each asset) we choose a portfolio (using a SSD strategy) 
on  $31^{\text{st}}$ December 2018, evaluate its performance out-of-sample for the next 20 business days so from $31^{\text{st}}$ December 2018 to $31^{\text{st}}$ January 2019, then repeat the process until the data is exhausted. In total this involved 60 out-of-sample periods for which we then have a single out-of-sample time series of returns. For simplicity we assume no transaction costs.

In order to define sector bounds, for a given sector $k$ we take its (fractional) exposure from Table~\ref{table1} as $\delta_k$ and define an interval $\Delta = 0.05$ such $\delta_k^L = (1 - \Delta) \delta_k $ and $\delta_k^U = (1 + \Delta) \delta_k$, where
$\delta_k^L$ and $\delta_k^U$ limit exposure to any particular sector, as in Equation~(\ref{eqjebdelta}).
These bounds apply to both subset SSD and standard SSD. 
Note here that these bounds preclude the possibility of any sector having zero investment.
This choice ensures that the portfolios chosen under both subset and standard SSD
have similar exposure to S\&P~500 sectors, whilst at the same time giving some leeway to the SSD optimiser in its choice of portfolio.



For numeric insight into the performance 
we show some selected comparative statistics in Table~\ref{table2}. These are calculated from the out-of-sample returns for the 
subset SSD approach, 
and correspondingly for the S\&P~500 index. 
For convenience this table also shows comparative statistics for standard SSD, but these will be discussed in the next section. 

Let $Q$ be a series of $0,\ldots,T$ daily portfolio values, where $Q_t$ is the value of the given portfolio on day $t$. 
In Table~\ref{table2} \textbf{FV} stands for the final portfolio value, assuming a starting amount of \$1, and is calculated as $Q_{T}/Q_{0}$. \textbf{CAGR} stands for Capital Annualised Growth Rate and as a percentage is calculated as $100  \left( \left(\frac{Q_T}{Q_0}\right)^{\frac{1}{Y}} -1 \right)$, where $Y = T/252$ is an approximation for the number of years in the out-of-sample period. \textbf{Sharpe} and \textbf{Sortino} are the annualised Sharpe and Sortino ratios respectively, where for their calculation we use the CBOE 10-year treasury notes (symbol TNX) as the risk-free rate.
 For  these four 
performance measures high values are better. 

\textbf{Vol} represents the annualised sample standard deviation of the out-of-sample returns. 
 \textbf{MDD} represents the maximum drawdown and as a percentage is calculated as $\max \left(0, 100 \max_{0 \leq t < u \leq T} \frac{Q_t - Q_u}{Q_t} \right)$. 
 For both Vol and MDD low values are better.

To evaluate diversification, Table~\ref{table2} shows two other statistics: across all rebalances, $\mu(C)$ is the average portfolio 
cardinality (number of assets in the portfolio chosen) and $\mu(W)$ is the average weight (as a percentage) over the assets chosen. Here $\mu(W) = 100/\mu(C)$ but we have given both of these statistics for completeness.


\begin{table}[!ht]
\centering
{\small
\renewcommand{\tabcolsep}{1mm} 
\renewcommand{\arraystretch}{1.4} 
\begin{tabular}{|l|rrrr|rr|rr|}
\hline

\multicolumn{1}{|c|}{Strategy} & \multicolumn{1}{c}{FV} & \multicolumn{1}{c}{CAGR} &  \multicolumn{1}{c}{Sharpe} & \multicolumn{1}{c}{Sortino} &
\multicolumn{1}{|c}{Vol} 
& \multicolumn{1}{c}{MDD} & \multicolumn{1}{|c}{$\mu(C)$} & \multicolumn{1}{c|}{$\mu(W)$}\\
\hline
Subset SSD (scaled)     &     1.93 &    14.02 &     0.66 &     0.91 &    19.22 &    31.85 &    65.37 &     1.53\\
Subset SSD (unscaled)   &     1.57 &     9.41 &     0.48 &     0.67 &    16.97 &    29.76 &    72.20 &     1.39\\
 \hline
 Standard SSD (scaled)   &     1.86 &    13.25 &     0.60 &     0.84 &    20.43 &    31.90 &    19.22 &     5.20\\
Standard SSD (unscaled) &     1.39 &     6.80 &     0.34 &     0.46 &    17.31 &    28.55 &    22.18 &     4.51\\
\hline
SP500                   &     1.90 &    13.74 &     0.61 &     0.84 &    21.31 &    33.92 & --       & --      \\
\hline

\end{tabular}
}
\caption{Comparative out-of-sample statistics}
\label{table2}
\end{table}

With regard to the scaling of tails,~\cite{fabian2011b, roman2013, valle2017} all concluded that scaled SSD tends to achieve superior out-of-sample returns, but not necessarily superior risk, when compared to unscaled SSD. The reason for this is that by scaling the tails more importance is given to the returns in the right tails of the distribution. 

In Table~\ref{table2} we observe the same behaviour, 
with the scaled version of subset SSD 
outperforming the unscaled version in terms of performance (FV, CAGR). The gain in absolute performance also translates to better risk-adjusted performance (Sharpe, Sortino).

 As can be seen from
Table~\ref{table2}
unscaled subset SSD does not  show superior performance with regard to  four of the six performance measures when compared to the S\&P~500. 
Scaled subset SSD outperformed 
the S\&P~500 for all six of the performance measures in Table~\ref{table2}. This is despite the substantive  Covid drop in 2020 that can be seen in Figure~\ref{fig1}.

\begin{figure}[!htb]
\centering
\includegraphics[width=1\textwidth]{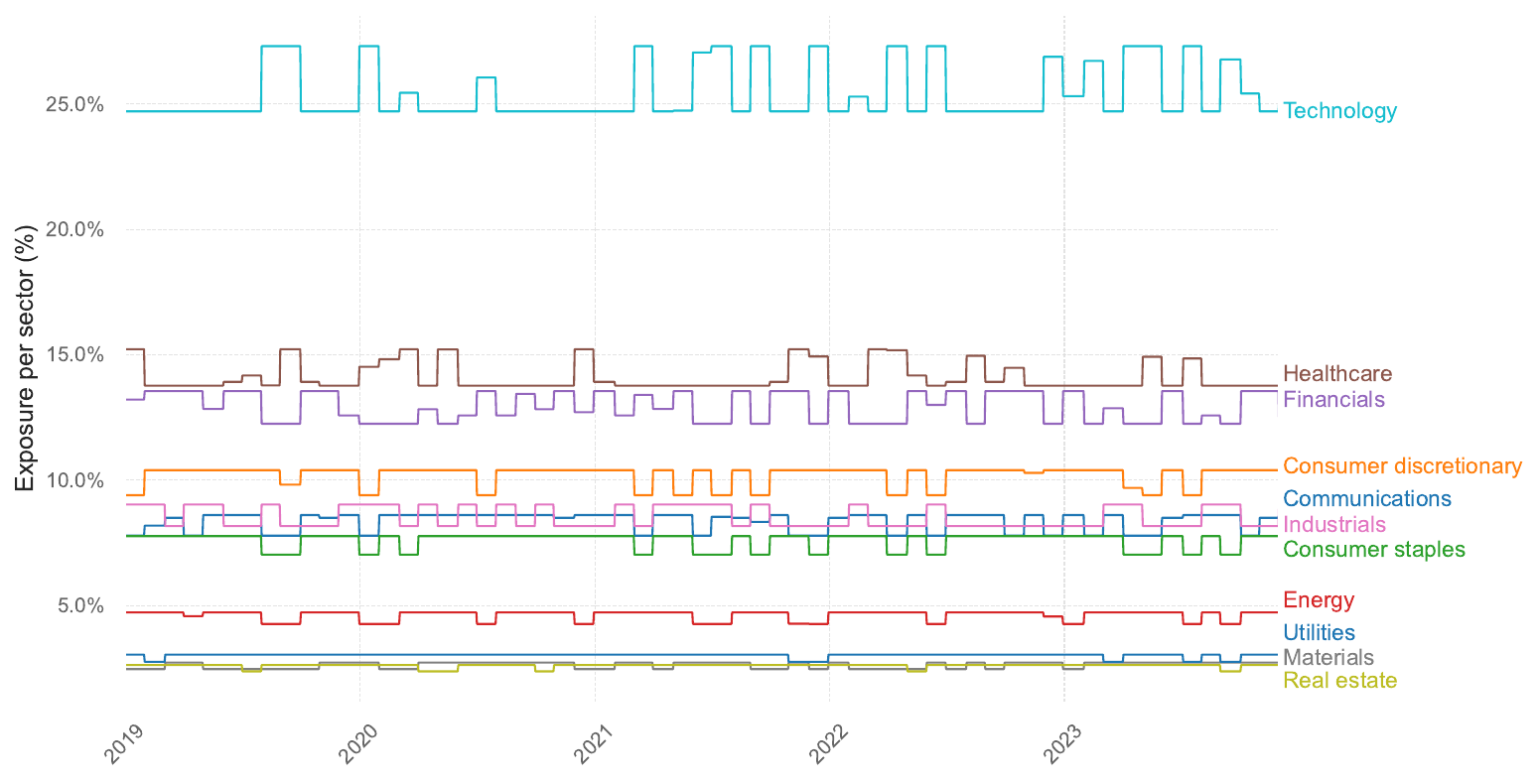}
  \caption{Out-of-sample exposure per sector, scaled subset SSD}
  \label{fig3}
\end{figure}

Figure~\ref{fig3} shows the out-of-sample exposure per sector for scaled subset SSD. Note here that this exposure is limited to
be within $\Delta$, here 5\%, of the 
S\&P~500
sector weightings shown in Table~\ref{table1}. Figure~\ref{fig3}  is designed to  highlight the rebalances where 
exposure changes. 
So we can see, for example, that at the start of 2020 there is an increase at a rebalance in the exposure to technology stocks (the top line in Figure~\ref{fig3}).
A short while later, at another rebalance, this exposure drops back to its previous level.
Note here that Figure~\ref{fig3} does not plot
exposure day by day as prices change and hence exposure naturally changes.

Despite the potentially exponential number of constraints involved in the cutting plane procedures required to solve  our subset SSD formulation to proven
optimality  our experience has been that 
the computational effort required  was small. 
In our experiments a total of 60 rebalances were needed. For scaled subset SSD, for example,  the average computational time per rebalance was 0.58s, with a maximum of 1.86s and a minimum of 0.13s (median 0.54s).
For  the other SSD cases in Table~\ref{table2} the average computational time was between 0.30s and 0.35s and no rebalance required more than a second.

\subsection{Out-of-sample performance - standard SSD with sector bounds}
\label{sec:standardout}

In this section we evaluate the performance of the standard SSD 
approach, which was outlined above in Section~\ref{secalt},
when compared to both our subset SSD approach and the S\&P~500.

We would remind the reader here that the main difference between subset SSD and the 
standard SSD approach  is that in subset SSD we actively try to find sector portfolios that SSD dominate their respective 
sector indices, as opposed to standard SSD where there is no attempt to ensure this.

Table~\ref{table2} shows the results obtained with standard SSD with sector bounds.
In that table we observe the same behaviour as for subset SSD in that
 the scaled version
outperforms the unscaled version in terms of performance (FV, CAGR). The gain in absolute performance also translates to better risk-adjusted performance (Sharpe, Sortino). 
Scaled standard SSD outperforms the S\&P~500 for only two of the six performance measures.

With regards to diversification, we have from 
Table~\ref{table2} that 
subset SSD  yielded portfolios with a much larger number of assets and hence a much smaller average weight than standard SSD, in both the scaled and unscaled versions. On average subset SSD had  over three times as many assets as standard SSD, both for the scaled and unscaled versions.

From Table~\ref{table2} and the above discussion it seems reasonable to  conclude that for the S\&P~500 data we examined:
\textbf{\emph{scaled subset SSD outperforms both standard SSD and the S\&P~500.}}

\subsection{Varying sector bounds}

To investigate the performance of our subset SSD approach when we varied sector bounds we performed twenty different 
experiments. As above,
in order to define sector bounds for a given sector $k$ we take its exposure from Table~\ref{table1} as $\delta_k$. Using
 $\Delta$ we have $\delta_k^L = (1 - \Delta) \delta_k $ and $\delta_k^U = (1 + \Delta) \delta_k$, where (as before)
$\delta_k^L$ and $\delta_k^U$ limit exposure to any particular sector, as in Equation~(\ref{eqjebdelta}).

We evaluated the  out-of-sample performance of both scaled subset SSD and scaled standard SSD for $\Delta = (0.01, 0.02, \dots, 0.20)$.
The results can be seen in Table~\ref{table3}. In this table we have, for example for FV and scaled subset SSD, that over the twenty 
values of $\Delta$ considered, the mean FV value was 1.93, the median FV value was 1.93, the minimum FV value was 1.91 and the maximum FV value was 1.94.

\begin{table}[!htb]
\centering
{\small
\renewcommand{\arraystretch}{1.5}
\begin{tabular}{|l|rrrr|rrrr|r|}
\hline
\multirow{2}{*}{Stats} & \multicolumn{4}{c|}{Subset SSD (scaled)} & \multicolumn{4}{c|}{Standard SSD (scaled)} & \multirow{2}{*}{ S\&P~500 } \\
\cline{2-9}
& Mean & Median & Min & Max & Mean & Median & Min & Max & \\
\hline
FV & 1.93 & 1.93 & 1.91 & 1.94 & 1.84 & 1.84 & 1.79 & 1.87 & 1.90 \\
CAGR & 14.03 & 14.05 & 13.83 & 14.18 & 12.97 & 13.04 & 12.33 & 13.33 & 13.74 \\
Sharpe & 0.67 & 0.66 & 0.66 & 0.67 & 0.59 & 0.59 & 0.56 & 0.61 & 0.61 \\
Sortino & 0.91 & 0.91 & 0.90 & 0.92 & 0.82 & 0.82 & 0.77 & 0.84 & 0.84 \\
\hline
Vol & 19.17 & 19.17 & 19.10 & 19.25 & 20.49 & 20.45 & 20.40 & 20.59 & 21.31 \\
MDD & 31.89 & 31.89 & 31.82 & 31.95 & 32.16 & 31.98 & 31.66 & 32.91 & 33.92 \\
 \hline
 $\mu(C)$ & 65.37 & 65.37 & 65.37 & 65.37 & 19.13 & 19.14 & 18.87 & 19.53 & -- \\
 $\mu(W)$ & 1.53 & 1.53 & 1.53 & 1.53 & 5.23 & 5.22 & 5.12 & 5.30 & -- \\
\hline

\end{tabular}
}
\caption{Summary statistics for the scaled formulations when $\Delta = (0.01, 0.02, \ldots, 0.20)$}
\label{table3}
\end{table}

Considering Table~\ref{table3} we can see that for
 the four 
performance measures where high values are better (so FV, CAGR, Sharpe and Sortino) the \emph{minimum} values for these measures for 
scaled subset SSD exceed the \emph{maximum} values for these measures for scaled standard SSD. For the two performance measures 
where low values are better (so Vol and MDD) the \emph{maximum} value for Vol for scaled subset SSD is below the 
\emph{minimum} value for Vol for scaled standard SSD. For MDD the \emph{maximum} value for Vol for scaled subset SSD is only slightly above  the 
\emph{minimum} value for Vol for scaled standard SSD.
Therefore varying the sector bounds, as summarised in Table~\ref{table3}, leads us to reasonably conclude that 
\textbf{\emph{scaled subset SSD is a better strategy to use than scaled standard SSD.}}


In a similar fashion
for the four 
performance measures where high values are better (so FV, CAGR, Sharpe and Sortino) the minimum values for these measures for 
scaled subset SSD exceed  the values associated with the S\&P~500.
 For the two performance measures 
where low values are better (so Vol and MDD) the maximum values for these measures for scaled subset SSD are 
below the values associated with the S\&P~500.
In other words \textbf{\emph{with regard to all six performance measures scaled subset SSD dominates the S\&P~500.}}

One feature of Table~\ref{table3} is that for subset SSD, over the twenty values of $\Delta$ considered,  the average portfolio cardinality
$\mu(C)$ and  average asset weight $\mu(W)$ never change.

In our S\&P~500 data our sectors are distinct, so we always considering just a single 
asset subset ($k$, say)  in Equations~(\ref{ajeb1})-(\ref{ajeb7}). As commented above
in Section~\ref{sec:gamma},  and proved in~\ref{append1},  this means that at a rebalance we find a SSD portfolio for  
subset $k$ and scale the weights in that SSD portfolio by $W^k_{opt}$, the weighting for that asset subset as decided by solving 
our subset SSD problem Equations~(\ref{exjebt4sz})-(\ref{zreal}).

Hence, ignoring degeneracy, over the set of rebalances for  asset subset $k$
we will always be finding the same sequence of SSD portfolios for that asset subset whatever the value 
of $\Delta$.
Consequently, over all asset subsets,  $\mu(C)$, 
the average portfolio cardinality, will be the same for every $\Delta$ (and so $\mu(W)=100/\mu(C)$ will also always be the same).

\subsection{Out-of-sample performance - no sector constraints}

Imposing sector constraints cannot increase the  in-sample performance of any portfolio. Rather it will (typically) decrease that performance as compared with the same approach but with no sector constraints.  Clearly what matters in practice is not in-sample performance, rather it is 
out-of-sample performance, i.e.~the performance of the in-sample decided portfolio once it is bought and held out-of-sample into the future. Imposing sector constraints
may lead to improved out-of-sample performance.

To investigate this issue for our S\&P~500 data we show in Table~\ref{table6} the out-of-sample performance 
for both subset SSD and standard SSD when we have no sector constraints imposed on the choice of in-sample portfolio. 
Having no sector constraints is equivalent to  setting
$\delta^L_k =0~\text{and}~\delta^U_k=1~ k=1,\ldots,K$ in 
Equation~(\ref{eqjebdelta}). 

Comparing the results in Table~\ref{table6} with the corresponding results in Table~\ref{table2} we have that for the four 
performance measures (FV, CAGR, Sharpe and Sortino) where higher values are better each of the four strategies in Table~\ref{table2} exhibit 
better performance than they do in Table~\ref{table6}. 
Across these four measures and four strategies each and every one of the 16 values in Table~\ref{table2}  are better than the corresponding values in Table~\ref{table6}.

With regard to Vol and MDD (where lower values are better) the performance is more 
mixed. Across these two measures and four strategies  half of the eight values are better in Table~\ref{table2}, half are better in Table~\ref{table6}. 

\textbf{\emph{On balance however it seems reasonable to conclude that, for the S\&P~500 data that we have examined, imposing sector constraints results in 
improved out-of-sample performance, irrespective of the strategy adopted.}}

\begin{table}[!ht] 
\centering
{\small
\renewcommand{\tabcolsep}{1mm} 
\renewcommand{\arraystretch}{1.4} 
\begin{tabular}{|l|rrrr|rr|rr|} 
\hline
\multicolumn{1}{|c|}{Strategy} & \multicolumn{1}{c}{FV} & \multicolumn{1}{c}{CAGR} &  \multicolumn{1}{c}{Sharpe} & \multicolumn{1}{c}{Sortino} &
\multicolumn{1}{|c}{Vol} 
& \multicolumn{1}{c}{MDD} & \multicolumn{1}{|c}{$\mu(C)$} & \multicolumn{1}{c|}{$\mu(W)$}\\
\hline
Subset SSD (scaled)     &     1.75 &    11.88 &     0.58 &     0.79 &    18.53 &    33.03 &    65.37 &     1.53\\
Subset SSD (unscaled)   &     1.38 &     6.65 &     0.33 &     0.45 &    17.31 &    33.20 &    72.20 &     1.39\\
\hline
Standard SSD (scaled)   &     1.82 &    12.69 &     0.54 &     0.77 &    22.66 &    31.05 &    11.78 &     8.49\\
Standard SSD (unscaled) &     1.25 &     4.60 &     0.23 &     0.32 &    14.26 &    22.18 &    16.63 &     6.01\\
\hline
SP500                   &     1.90 &    13.74 &     0.61 &     0.84 &    21.31 &    33.92 & --       & --      \\
\hline

\end{tabular} } 
\caption{Comparative out-of-sample statistics, no sector constraints} 
\label{table6} 
\end{table}

\subsection{Fama-French out-of-sample performance}

In this section we evaluate our approach when the asset universe is comprised of the 
Fama-French 49 industry portfolios. We employed two benchmarks: an equally-weighted index 
calculated daily from all portfolios and the CRSP NYSE value-weighted index. As sector indices we employed both the 
equally-weighted indices of each sector as well as the CRSP sector indices mentioned previously.

We used the same setup as in Section~\ref{sec:subsetout} 
and Section~\ref{sec:standardout}: 60 in-sample returns with rebalancing every 21 days, no 
transaction costs and $\Delta = 0.05$. Computation times were even smaller than those reported
 above with regard to the S\&P~500
since the asset universe comprises far fewer assets. The results with  equally-weighted indices are shown in Table~\ref{table4}, and the
 results with the CRSP indices are shown in Table~\ref{table5}.

With regard to Table~\ref{table4} we have that scaled subset SSD outperforms unscaled subset SSD across all six of the performance 
measures shown. Scaled subset SSD is superior to scaled standard SSD with 
respect to four measures (Sharpe, Sortino, Vol and MDD), inferior with respect to FV and CAGR. Both scaled subset SSD and scaled 
standard SSD outperformed the equally-weighted index with respect to all six performance measures. As we observed previously with 
our S\&P~500 results subset SSD portfolios are much more diversified than their standard SSD counterparts.

\begin{table}[!ht]
\centering
{\small
\renewcommand{\tabcolsep}{1mm} 
\renewcommand{\arraystretch}{1.4} 
\begin{tabular}{|l|rrrr|rr|rr|} 
\hline
\multicolumn{1}{|c|}{Strategy} & \multicolumn{1}{c}{FV} & \multicolumn{1}{c}{CAGR} &  \multicolumn{1}{c}{Sharpe} & \multicolumn{1}{c}{Sortino} &
\multicolumn{1}{|c}{Vol} 
& \multicolumn{1}{c}{MDD} & \multicolumn{1}{|c}{$\mu(C)$} & \multicolumn{1}{c|}{$\mu(W)$}\\
\hline
Subset SSD (scaled)    &     2.08 &    15.83 &     0.83 &     1.14 &    20.25 &    34.80 &    27.45 &     3.64\\
Subset SSD (unscaled) &     1.97 &    14.55 &     0.77 &     1.07 &    20.31 &    35.67 &    28.78 &     3.47\\
\hline
Standard SSD (scaled)    &     2.11 &    16.16 &     0.82 &     1.10 &    21.12 &    37.43 &    12.52 &     7.99\\
Standard SSD (unscaled) &     1.72 &    11.44 &     0.64 &     0.87 &    19.97 &    36.87 &    12.42 &     8.05\\
\hline
Equally-weighted               &     2.02 &    15.16 &     0.75 &     1.04 &    22.30 &    38.33 & --       & --      \\
\hline
\end{tabular}
}
\caption{Comparative out-of-sample statistics, Fama-French with equally-weighted indices}
\label{table4}
\end{table}

With regard to Table~\ref{table5} and the CRSP indices we have that scaled subset SSD performs better than
unscaled subset SSD (superior in four performance measures, inferior in two).
Scaled subset SSD outperformed scaled standard SSD for five of the six performance measures.
Both subset SSD approaches achieved better performance (in all six performance measures)  as compared with the 
CRSP NYSE  value-weighted index.

\begin{table}[!ht]
\centering
{\small
\renewcommand{\tabcolsep}{1mm} 
\renewcommand{\arraystretch}{1.4} 
\begin{tabular}{|l|rrrr|rr|rr|} 
\hline
\multicolumn{1}{|c|}{Strategy} & \multicolumn{1}{c}{FV} & \multicolumn{1}{c}{CAGR} &  \multicolumn{1}{c}{Sharpe} & \multicolumn{1}{c}{Sortino} &
\multicolumn{1}{|c}{Vol} 
& \multicolumn{1}{c}{MDD} & \multicolumn{1}{|c}{$\mu(C)$} & \multicolumn{1}{c|}{$\mu(W)$}\\
\hline
Subset SSD (scaled)     &     2.07 &    15.66 &     0.82 &     1.13 &    20.22 &    35.42 &    24.88 &     4.02\\
Subset SSD (unscaled)  &     2.01 &    15.05 &     0.80 &     1.10 &    20.09 &    34.97 &    27.12 &     3.69\\
\hline
Standard SSD (scaled)  &     1.97 &    14.59 &     0.78 &     1.06 &    20.15 &    36.15 &    13.17 &     7.59\\
Standard SSD (unscaled) &     1.67 &    10.88 &     0.62 &     0.84 &    19.74 &    37.03 &    12.60 &     7.94\\
\hline
CRSP NYSE value-weighted       &     1.73 &    11.65 &     0.64 &     0.88 &    20.76 &    38.37 & --       & --      \\
\hline
\end{tabular}
}
\caption{Comparative out-of-sample statistics, Fama-French with CRSP indices}
\label{table5}
\end{table}

On balance the results for Fama-French data in this section  confirm the effectiveness of 
our (scaled) subset SSD approach as compared to other approaches.

\section{Conclusions}
\label{sec6}

In this paper we have considered the problem of
how to construct a portfolio that is
designed to outperform a given market index, whilst having regard to the proportion of 
the portfolio invested in distinct market sectors (asset subsets).

We presented a new approach, subset SSD, for the problem.
Subset SSD explicitly takes into account the performance of an index for each sector
as well as the performance of the market index.
In our approach portfolios associated with each sector are treated in a SSD manner so that we 
actively try to find sector portfolios that SSD dominate their respective sector indices.

Computational results were given for our subset SSD approach as applied to the S\&P~500 over the period 
$3^{\text{rd}}$ October 2018 to $29^{\text{th}}$ December 2023, with this data 
being made publicly available for the benefit of future researchers.
Our subset SSD approach requires the solution of a multivariate second-order stochastic dominance problem, but for the data we considered
the computation time for this was negligible. 

Our computational results indicated that the scaled version of our subset SSD
approach  outperforms the S\&P~500 over the period considered. Our approach also outperforms the standard SSD based
approach to the problem. Our results show, that for the S\&P~500 data we considered, 
 including sector constraints improves out-of-sample performance, irrespective of the SSD approach adopted.
Results given for Fama-French data involving 49 industry portfolios also confirmed the effectiveness of our subset SSD approach.


\appendix
\section{Cutting plane procedure for subset SSD}
\label{append}
The  cutting plane solution procedure for
our subset SSD formulation, Equations~(\ref{exjebt4sz})-(\ref{zreal}), is as given below.


First define an initial scenario set $\mathcal{J^*}$ where there is at least one set of cardinality $s$
in $\mathcal{J^*}$
for all values of $s=1,\ldots,S$ .
We have one scenario for each in-sample time period considered, where a single scenario
consists of  the return for each  asset, as well as the overall market index return and the index return for 
each asset subset.
Amend Equation~(\ref{exjebt5slina}) to 
\begin{equation}
\mathcal{Z}_s^k \leq \frac{1}{S} \sum_{j \in \mathcal{J}} \sum_{i \in N^k} r_{ij} w_i – W^k \hat{\tau}_s^k~~~~\forall \mathcal{J} \in \mathcal{J^*},~|\mathcal{J}| = s,~s=1,\ldots,S,~k=0,\ldots,K
\label{exjebt6slina}
\end{equation}

\begin{enumerate}
\item Solve the amended optimisation program, optimise Equation~(\ref{exjebt4sz}) subject to 
Equations~(\ref{exjebt5slina1})-(\ref{zreal}),(\ref{exjebt6slina})

\item Consider all values of $s$ and $k$ ($s=1,\ldots,S,~k=0,\ldots,K$) in turn and if in the solution to the amended optimisation program 
\begin{equation}
\mathcal{Z}_s^k >\frac{1}{S} (\mbox{sum of the $s$ smallest portfolio returns in subset $k$ over the $S$ scenarios)} – W^k \hat{\tau}_s^k
\label{jebt9az}
\end{equation}

\noindent then add the scenario set associated with these 
$s$ smallest returns to $\mathcal{J^*}$. Here the scenario set that is added constitutes a valid cut associated
with Equation~(\ref{exjebt5slina}) that is violated by the current solution. 
\item If scenarios sets have been added to $\mathcal{J^*}$ go to Step (1), else terminate with the optimal solution
to the original problem, Equations~(\ref{exjebt4sz})-(\ref{zreal}).
\end{enumerate}

\section{SSD equivalence – single asset subset}
\label{append1}

If we have just a single asset subset in $\Gamma$ in Equations~(\ref{ajeb1})-(\ref{ajeb7}) then (assuming this is asset subset one for convenience) that optimisation becomes
 
\begin{equation}
\mbox{maximise}~\mathcal{V}
\label{ajeb1a}
\end{equation}
subject to 
\begin{equation}
\mathcal{V}_s^1 \leq \frac{1}{S} \sum_{j \in \mathcal{J}} \sum_{i \in N^1} r_{ij} w_i /W^1_{opt}- \hat{\tau}_s^1~~~~\forall \mathcal{J} \subseteq \{1, ..., S\},~|\mathcal{J}| = s,~s=1,\ldots,S
\label{ajeb2a1}
\end{equation}
\begin{equation}
\beta_s \mathcal{V} \leq W^1_{opt} \mathcal{V}_s^1 ~~~~s=1,\ldots,S
\label{ajeb2aa}
\end{equation}
\begin{equation}
W^1_{opt} = \sum_{i \in N^1} w_i
\label{ajeb3a}
\end{equation}
\begin{equation}
 w_i \geq 0~~~~\forall i \in N^1
\label{ajeb5a}
\end{equation}
\begin{equation}
\mathcal{V} \in\mathbb{R}
\label{ajeb6a}
\end{equation}
\begin{equation}
\mathcal{V}_s^1 \in\mathbb{R}~~~~s=1,\ldots,S
\label{ajeb7a}
\end{equation}

To see that this optimisation reduces to a standard SSD optimisation, let $x_i =w_i / W^1_{opt}$ and  $Y= \mathcal{V}/W^1_{opt}$. Then the above optimisation becomes:

\begin{equation}
\mbox{maximise}~W^1_{opt}Y
\label{ajeb1ab}
\end{equation}
subject to 
\begin{equation}
\mathcal{V}_s^1 \leq \frac{1}{S} \sum_{j \in \mathcal{J}} \sum_{i \in N^1} r_{ij} x_i - \hat{\tau}_s^1~~~~\forall \mathcal{J} \subseteq \{1, ..., S\},~|\mathcal{J}| = s,~s=1,\ldots,S
\label{ajeb2ab}
\end{equation}
\begin{equation}
\beta_s Y \leq  \mathcal{V}_s^1 ~~~~s=1,\ldots,S
\label{ajeb2aab}
\end{equation}
\begin{equation}
\sum_{i \in N^1} x_i = 1
\label{ajeb3ab}
\end{equation}
\begin{equation}
 x_i \geq 0~~~~\forall i \in N^1
\label{ajeb5ab}
\end{equation}
\begin{equation}
Y \in\mathbb{R}
\label{ajeb6ab}
\end{equation}
\begin{equation}
\mathcal{V}_s^1 \in\mathbb{R}~~~~s=1,\ldots,S
\label{ajeb7ab}
\end{equation}
Noting that in the objective function, Equation~(\ref{ajeb1ab}) the $W^1_{opt}$ term is a constant that can be dropped we 
have that Equations~(\ref{ajeb1ab})-(\ref{ajeb7ab}) is the problem of finding a SSD portfolio for the assets in subset $ N^1$.

In other words if we have just a single asset subset to consider the original optimisation problem, 
Equations~(\ref{ajeb1})-(\ref{ajeb7}), is equivalent to finding a SSD portfolio for the asset subset.
The weights $w_i$ associated with the assets in the subset considered are equal to  $ W^1_{opt} x_i$. So 
the $ W^1_{opt}$ acts as  a scaling factor applied to the weights $x_i$ in the subset SSD portfolio to give the
optimal  
weights $w_i$ for the original optimisation problem.

\bibliographystyle{plainnat}
\bibliography{paper}

\end{document}